\documentclass[twocolumn,preprintnumbers,superscriptaddress,landscape,nofootinbib]{revtex4}

\preprint{RBRC-1322}

\usepackage{amsfonts}
\usepackage{amsmath}
\usepackage{latexsym}
\usepackage{xcolor}
\usepackage{graphicx}
\usepackage{amssymb}
\usepackage{epsfig}
\usepackage{epstopdf}

\usepackage{qcircuit}

\allowdisplaybreaks
\usepackage[colorlinks=true, pdfstartview=FitV, linkcolor=red, citecolor=blue, urlcolor=blue]{hyperref}

\def\del{\partial}

\renewcommand{\thefootnote}{\fnsymbol{footnote}}

\renewcommand{\bar}{\overline}
\newcommand{\br}{\notag\\&\;\;\;\;\;\;}

\newcommand{\NN} {\nonumber}

\newcommand{\p}{\partial}

\newcommand{\im}{\mathrm{i}}
\newcommand{\e}{\mathrm{e}}

\newcommand{\calO}{\mathcal{O}}

\newcommand{\mA}{\mathrm{A}}

\begin{document}

\date{\today}
\title{Classically Emulated Digital Quantum Simulation of the Schwinger Model\\ 
with Topological Term via Adiabatic State Preparation}

\author{Bipasha Chakraborty}
\email[]{bc335ATdamtp.cam.ac.uk}
\affiliation{{\it
Department of Applied Mathematics and Theoretical Physics,
Centre for Mathematical Sciences, Wilberforce Road, Cambridge, CB3 0WA, UK
}}

\author{Masazumi Honda}
\email[]{masazumi.hondaATyukawa.kyoto-u.ac.jp}
\affiliation{{\it
Department of Applied Mathematics and Theoretical Physics,
Centre for Mathematical Sciences, Wilberforce Road, Cambridge, CB3 0WA, UK
}}

\author{Taku Izubuchi}
\email[]{izubuchiATquark.phy.bnl.gov}
\affiliation{{\it
Physics Department, Brookhaven National Laboratory, Upton, New York 11973, USA
}}
\affiliation{{\it
RIKEN BNL Research center, Brookhaven National Laboratory, Upton, NY, 11973, USA
}}

\author{Yuta Kikuchi}
\email[]{yuta.kikuchiATriken.jp}
\affiliation{{\it
RIKEN BNL Research center, Brookhaven National Laboratory, Upton, NY, 11973, USA
}}

\author{Akio Tomiya}
\email[]{akio.tomiyaATriken.jp}
\affiliation{{\it
RIKEN BNL Research center, Brookhaven National Laboratory, Upton, NY, 11973, USA
}}

\begin{abstract}
We designed a protocol for digital quantum computation of a gauge theory with a topological term in Minkowski spacetime, 
which is practically inaccessible by standard lattice Monte Carlo simulations. 
We focus on $1+1$ dimensional quantum electrodynamics with the $\theta$-term 
known as the Schwinger model and test our protocol for this on IBM simulator.
We construct the true vacuum state of a lattice Schwinger model using adiabatic state preparation
which, in turn, allows us to compute an expectation value of the fermion mass operator with respect to the vacuum. 
Upon taking a continuum limit
we find that our result in massless case agrees with the known exact result. 
In massive case, we find an agreement with mass perturbation theory in small mass regime
and deviations in large mass regime.
We estimate computational costs required to take a reasonable continuum limit.
Our results imply that 
digital quantum simulation appears promising tool to explore 
non-perturbative aspects of gauge theories with real time and topological terms.
\end{abstract}
\maketitle

\renewcommand{\thefootnote}{\arabic{footnote}}
\setcounter{footnote}{0}

\setcounter{page}{1}

\section{Introduction}
Gauge theory plays a central role in understanding our universe 
as all the known fundamental forces are described in the framework of gauge theory.
Quantum Chromodynamics (QCD) is the gauge theory describing the strong interaction 
among quarks and gluons. 
Since QCD is asymptotically free,
we need a non-perturbative treatment at low-energy 
to interpolate perturbative picture of quarks/gluons and physics of hadrons.
The only successful first-principle approach to handle this is lattice QCD 
in which we conventionally consider QCD on 4d Euclidean spacetime 
and discretize the spacetime by lattice to make the path integral finite dimensional. 
Evaluating the regularized path integral numerically and taking a continuum limit carefully,
one can study non-perturbative phenomena such as confinement and chiral symmetry breaking, and reproduce correct hadron spectrum \cite{Ratti:2018ksb, Lin:2011ti, Philipsen:2012nu, Ding:2015ona}.
Numerical integration of path integral in lattice field theory is usually done by the Markov-Chain Monte Carlo method which regards Boltzmann weight as a probability. 
Therefore it encounters a problem when the integrand is non-real positive and highly oscillating so that sampling becomes much less efficient. 
This problem known as the sign problem physically happens \textit{e.g.}~when we have topological terms ~\cite{Izubuchi:2008mu}, chemical potentials~\cite{Aarts:2015tyj} 
or real time
\cite{Takeda:2019idb}.
All of the above cases are crucial to understand our universe and therefore, an efficient way to explore the above situations is highly demanded \cite{Izubuchi:2008mu, Philipsen:2012nu, Ding:2015ona, Aarts:2015tyj, Takeda:2019idb}.

There are various approaches to challenge the sign problem within the framework of path integral formalism \cite{Aarts:2015tyj}; however, with limited success as the sign problem gets stronger. 
One may wonder if it is possible to attack gauge theories with the sign problem by switching to Hamiltonian formalism where the sign problem is absent from the beginning.
Instead, we have to regularize infinite dimensional Hilbert space
and play with huge vector space
whose dimension is typically more than exponential of spatial volume
in the unit of UV cutoff.
It seems beyond the capacity of current/near-future supercomputers
when spacetime dimension is not low.
However, it is reasonable to expect that
{\it quantum computers} do this job in the not-so-distant-future.
Anticipating growth of quantum computational resources,
it is worth to develop
methods to analyze gauge theories
suitable for quantum computers to prepare for the coming era of quantum supremacy \cite{Preskill:2012tg,Arute:2019zxq}.
It is particularly important to identify suitable algorithms and estimate computational resources required to take a reasonable continuum limit.

In this paper, we design a protocol 
for a {\it digital quantum simulation} of a gauge theory with a topological term on Minkowski spacetime 
which is practically inaccessible by the standard Monte Carlo approach.
We focus on the Schwinger model with the $\theta$-term~\cite{Schwinger:1962tn,Coleman:1975pw,Coleman:1976uz,Manton:1985jm},
which is $1+1$ dimensional $U(1)$ gauge theory coupled to a Dirac fermion 
described by the Lagrangian,
\begin{align}
\mathcal{L}_0
&= -\frac{1}{4}F_{\mu\nu}F^{\mu\nu} +\frac{g\theta}{4\pi}\epsilon_{\mu\nu}F^{\mu\nu} \br
+\im \bar{\psi} \gamma^\mu (\del_\mu +\im g A_\mu )\psi -m\bar{\psi}\psi ,
\label{eq:naiveL}
\end{align}
where $\gamma^0 =\sigma^3$, $\gamma^1 =\im\sigma^2$, $\gamma^5 = \gamma^0 \gamma^1$
and $F_{\mu\nu} = \del_\mu A_\nu -\del_\nu A_\mu$.
The physical parameters of this model are 
the gauge coupling $g$, topological angle $\theta$, and fermion mass $m$.
We discretize the space by lattice keeping time continuous 
and work in Hamiltonian formalism. 
Then we construct the true vacuum of the lattice Schwinger model at finite $(g,\theta ,m)$ by a digital quantum simulation via adiabatic state preparation
and compute the vacuum expectation value (VEV) of the fermion mass operator $\bar{\psi}\psi$. 
We take a continuum limit and
find that our result in massless case agrees 
with the exact result known in literature \cite{Hetrick:1988yg,Hetrick:1995wq,Hosotani:1996hy,Hosotani:1996sn}.
In massive case, we find an agreement with mass perturbation theory \cite{Adam:1998tw,Adam:1997wt} for small $m$ 
and deviations for large $m$.
Our results imply that 
digital quantum simulation is already useful tool to explore 
non-perturbative aspects of gauge theories with topological terms on Minkowski spacetime
even in current computational resource.
Here we use a classical simulator for quantum hardware rather than real quantum computers
for the purpose of designing quantum algorithms for gauge theories.
Its implementations on a real quantum device is left as a future work, that is another vital task especially in the forthcoming Noisy Intermediate-Scale Quantum (NISQ) era~\cite{Preskill2018quantumcomputingin}.

Many efforts have already been made on designing and implementing digital quantum simulations of quantum field theories~\cite{Jordan:2011ne,Jordan:2011ci,Jordan:2014tma,Garcia-Alvarez:2014uda,Wiese:2014rla,Marcos:2014lda,Mezzacapo:2015bra,Martinez:2016yna,Muschik:2016tws,Macridin:2018gdw,Lamm:2018siq,Klco:2018zqz,Klco:2018kyo,Gustafson:2019mpk,Alexandru:2019ozf,Klco:2019xro,Klco:2019evd,Lamm:2019uyc,Magnifico:2019kyj,Mueller:2019qqj,Gustafson:2019vsd} 
as well as analog quantum simulations~\cite{Zohar:2012ay,Banerjee:2012pg,Zohar:2012xf,Banerjee:2012xg,Wiese:2013uua,Zohar:2015hwa,Bazavov:2015kka,Zohar:2016iic,Bermudez:2017yrq,Nature2017,Zache:2018jbt,Zhang:2018ufj,Lu:2018pjk,Surace:2019dtp}.
In particular, the Schwinger model provides an ideal laboratory 
for developing quantum algorithms with limited quantum resources foreseeing larger-scale digital quantum simulations of various gauge theories.
So far, 
applications of quantum algorithms for the Schwinger model are limited to $\theta =0$, and performed with a free vacuum and quenching evolution~\cite{Martinez:2016yna,Muschik:2016tws,Klco:2018kyo,Kokail:2018eiw,Magnifico:2019kyj}
while analogue quantum simulations have been implemented in \cite{Nature2017,Surace:2019dtp}. 

The present work demonstrates how to construct the true vacuum in an interacting gauge theory with the topological term by a digital quantum simulation. 
We believe that our results open up potential applications of digital quantum simulation to quantum field theory since the preparation of true ground state is indispensable to calculate various observables such as scattering amplitudes non-perturbatively\footnote{
Schwinger model with the $\theta$-term has been studied by other approaches 
without using quantum computing in \cite{Byrnes:2002gj,Buyens:2017crb,Kuramashi:2019cgs,Funcke:2019zna,Zache:2018cqq}.
}.

\section{Schwinger model as qubits}
\label{sec:qubits}
First, we rewrite the lattice Schwinger model in terms of spin operators which act on the Hilbert space represented by qubits according to \cite{Hamer:1997dx}.
Instead of directly analyzing the system with the Lagrangian \eqref{eq:naiveL},
we consider the Lagrangian obtained by the chiral rotation $\psi \rightarrow e^{\im\frac{\theta}{2}\gamma_5}\psi$ to absorb the $\theta$-term 
via the transformation of the path integral measure \cite{Fujikawa:1979ay}.
Therefore, 
we can study the same physics by the Lagrangian,
\begin{align}
\mathcal{L}
= -\frac{1}{4}F_{\mu\nu}F^{\mu\nu} +\im\bar{\psi} \gamma^\mu (\del_\mu +\im g A_\mu )\psi -m\bar{\psi}e^{\im\theta\gamma^5}\psi .
\end{align}
In the temporal gauge $A_0 =0$, 
the Hamiltonian of this model is
\begin{align}
H
=\int dx \Biggl[ -\im\bar{\psi} \gamma^1 ( \del_1 +\im gA_1 )\psi  +m\bar{\psi}e^{\im\theta\gamma^5}\psi +\frac{1}{2}\Pi^2
\Biggr],
\end{align}
where $\Pi \equiv \dot{A}^1$ is the conjugate momentum of $A^1$.
The gauge invariance of physical Hilbert space is guaranteed by imposing the Gauss law:
\begin{align}
0
=-\del_1 \Pi - g\bar{\psi}\gamma^0 \psi.
\end{align}

\subsection{Lattice theory with Staggered fermion}
To implement a quantum simulation algorithm,
we need a regularization to make the Hilbert space finite dimensional.
For the Schwinger model, this is done just by placing the theory on a lattice and imposing the Gauss law\footnote{
This is true for open boundary condition
while there is a remaining gauge degree of freedom for periodic boundary condition.
} \cite{Muschik:2016tws}. 
Let us consider the theory on 1d spatial lattice with $N$ sites and lattice spacing $a$ keeping the time continuous.
Using the staggered fermion \cite{Kogut:1974ag,Susskind:1976jm}, the lattice Hamiltonian is given by\footnote{
Note that staggered fermion in $1+1$ dimensions in Hamilton formalism has only one taste.
}
\begin{align}
H
&= 
-\im \sum_{n=1}^{N-1}\left( w-(-1)^n \frac{m}{2}\sin{\theta}\right)
 \Bigl[ \chi_n^\dag e^{\im\phi_n} \chi_{n+1} -{\rm h.c.} \Bigr] 
\br 
+m \cos{\theta}\sum_{n=1}^N (-1)^n \chi_n^\dag \chi_n 
+J \sum_{n=1}^{N-1} L_n^2  ,
\end{align}
where $w=1/(2a)$ and $J=g^2 a /2$.
We have rescaled the gauge operators 
as $\phi_n$ $\leftrightarrow$ $-ag A^1 (x)$ 
and $L_n$ $\leftrightarrow$ $-\Pi(x)/g$,
where $\phi_n$ lives on a site $n$ while $L_n$ lives on a link between sites $n$ and $n+1$. 
A two-component Dirac fermion $\psi(x)=\big(\psi_u(x),\psi_d(x) \big)^\mathsf{T}$ 
is translated into a pair of neighboring one-component fermions 
according to the correspondence (see Supplementary Material for details):
\begin{align}
\frac{\chi_n}{\sqrt{a}} \leftrightarrow \left\{ \begin{matrix} \psi_u (x) & n:{\rm even} \cr
\psi_d (x) & n:{\rm odd} \end{matrix}\right. .
\end{align}
They satisfy the (anti-)commutation relations
\begin{align}
\{ \chi_n^\dag ,\chi_m \} = \delta_{mn},\ \ \{ \chi_n ,\chi_m \} =0 ,\ \  
[\phi_n ,L_m] =\im\delta_{mn} ,
\label{eq:commutation}
\end{align}
and the Gauss law on the lattice is 
\begin{align}
L_n -L_{n-1} = \chi_n^\dag \chi_n -\frac{ 1-(-1)^n}{2}.
\label{eq:GaussL}
\end{align}

\subsection{Mapping to spin system}
We rewrite the system in terms of spin variables in three steps.
Firstly, we perform the Jordan-Wigner transformation \cite{Jordan1928},
which maps the fermions to spin variables as
\begin{align}
\chi_n =  \Bigl(  \prod_{\ell <n }- \im Z_\ell \Bigr) \frac{X_n -\im Y_n}{2} ,
\label{eq:Jordan}
\end{align}
where $(X_n ,Y_n, Z_n )$ stands for the Pauli matrices $(\sigma^1 ,\sigma^2 ,\sigma^3 )$ at site $n$. 
Secondly, we specify a boundary condition and solve the Gauss law.
We impose an open boundary condition 
which restricts $L_n$ to a constant at the boundary.
Solving the Gauss law,
we rewrite $L_n$ in terms of the spin variables as
\begin{align}
L_n  =L_0 + \frac{1}{2}\sum_{\ell =1}^n \left( Z_\ell +(-1)^\ell \right) ,
\end{align}
where the constant $L_0$ specifies our boundary condition.
The Schwinger model with $(\theta ,L_0 )$ is equivalent to the one with $(\theta +2\pi L_0 ,0 )$ \cite{Coleman:1976uz}
and therefore we can take $L_0 =0$ without loss of generality. 
Finally, 
we can eliminate $\phi_n$ by the redefinition\footnote{
If we took a periodic boundary condition,
then $L_0$ was dynamical and one of $\phi_n$'s could not be eliminated by the redefinition.
} 
$\chi_n $ $\rightarrow$ $\prod_{\ell <n} \left[ e^{-\im\phi_\ell} \right] \chi_n$.

Thus, the lattice Schwinger model is purely described in terms of the spin variables:
\begin{align}
\label{eq:spinHam}
H = H_{ZZ} +H_\pm +H_Z,
\end{align}
where
\begin{align}
\label{eq:Hspin_terms}
H_{ZZ} 
&= \frac{J}{2} \sum_{n=2}^{N-1}\sum_{1\leq k <\ell \leq n}  Z_k Z_\ell ,\NN\\
H_\pm 
&=  \frac{1}{2}\sum_{n=1}^{N-1}\left( w -(-1)^n \frac{m}{2}\sin{\theta} \right) \Bigl[X_nX_{n+1} +Y_nY_{n+1} \Bigr]  ,\NN\\
H_Z 
&=  \frac{m\cos{\theta}}{2} \sum_{n=1}^N (-1)^n  Z_n
 -\frac{J}{2} \sum_{n=1}^{N-1} (n \text{ mod } 2) \sum_{\ell =1}^n  Z_{\ell} ,
\end{align}
up to irrelevant constant terms.
Note that the nonlocal interactions in $H_{ZZ}$ show up as a consequence of solving the Gauss law constraint.
For the formulation based on \eqref{eq:naiveL}, see Supplementary Material.

\section{Adiabatic preparation of vacuum}
\label{sec:protocol}
We study the VEV of the mass operator:
\begin{align}
\langle \bar{\psi}(x) \psi (x) \rangle 
= \langle {\rm vac}| \bar{\psi}(x) \psi (x) |{\rm vac}\rangle ,
\end{align}
where $|{\rm vac}\rangle$ is the ground state of the full Hamiltonian $H$.
Here, instead of directly studying the local operator $\bar{\psi}(x) \psi (x)$,
we analyze the operator averaged over space:
\begin{align}
\frac{1}{2Na} \langle {\rm vac}|  \sum_{n=1}^{N} (-1)^n Z_n  |{\rm vac}\rangle  ,
\end{align}
whose continuum limit is 
the same as $\langle \bar{\psi}(x) \psi (x) \rangle $ by translational invariance.

We prepare the vacuum state $|{\rm vac}\rangle$ using the adiabatic theorem as follows.
We first choose an initial Hamiltonian 
$H_0$ of a simple system such that its ground state $|{\rm vac}\rangle_0$ 
is unique and known.
Next, we consider the following time evolution of $|{\rm vac}\rangle_0$:
\begin{align}
\mathcal{T}\exp{\left(-\im \int_0^T dt\ H_\mA (t) \right)} |{\rm vac}\rangle_0 ,
\end{align}
where $\mathcal{T}\exp$ denotes a time-ordered exponential.
The adiabatic Hamiltonian $H_\mA(t)$ is an hermitian operator satisfying
\begin{align}
\label{eq:adiabatic_cond}
H_\mA(0) = H_0, \quad H_\mA(T)=H.
\end{align}
The adiabatic theorem claims that, 
if the system described by the Hamiltonian $H_\mA (t)$ is gapped and has a unique ground state, 
then the ground state of $H$ is obtained by the time evolution
\begin{align}
|{\rm vac}\rangle = \lim_{T\rightarrow\infty} \mathcal{T}\exp{\left(-\im \int_0^T dt\ H_\mA (t) \right)} |{\rm vac}\rangle_0 .
\label{eq:evolution}
\end{align}
In practice, we take finite $T$ and discretize the integral,
and therefore we can obtain only an approximation of the vacuum. 
This implies that an expectation value of an operator under the approximate vacuum
has intrinsic systematic errors. 
In Supplementary Material, we discuss how we estimate the systematic errors.

In our simulation, 
we take the initial Hamiltonian $H_0$ as 
\begin{align}
H_0 = H_{\rm ZZ} +\left. H_{\rm Z} \right|_{m\rightarrow m_0 ,\theta\rightarrow 0} ,
\label{eq:initial}
\end{align}
where $m_0 \in\mathbb{R}_{\geq 0}$ can be arbitrary in principle, however, is chosen so that systematic errors become small.
The ground state of $H_0$ is
\begin{align}
 |{\rm vac}\rangle_0
=|0\rangle \otimes |1\rangle \otimes \cdots \otimes |0\rangle \otimes |1\rangle ,
\end{align}
where $Z|0\rangle =|0\rangle$ and $Z|1\rangle =-|1\rangle$.
In order to evolve it into the desired vacuum state we choose the following adiabatic Hamiltonian,
\begin{align}
\label{eq:adiabaticH}
H_\mA(t)=H_{ZZ} +H_{\pm,\mA}(t) +H_{Z,\mA}(t),
\end{align}
with $H_{\pm,\mA}$ and $H_{Z,\mA}$ obtained by replacing the parameters of $H_\pm$ and $H_Z$ in the Hamiltonian~\eqref{eq:spinHam} as
\begin{align}
\label{eq:ad_replace}
w \to \frac{t}{T}w, \ \ \theta\to \frac{t}{T}\theta,\ \ 
m\to \left( 1-\frac{t}{T}\right) m_0 +\frac{t}{T}m .
\end{align}
We take finite $T$ and approximate the time evolution \eqref{eq:evolution} by \cite{Lloyd1073,Suzuki1991}
\begin{align}
|{\rm vac}\rangle 
\simeq  U(T )U(T-\delta t ) \cdots U(2\delta t ) U(\delta t )  |{\rm vac}\rangle_0 ,
\label{eq:evolution_ST}
\end{align}
where  $U(t) =e^{-\im H_A (t)\delta t}$ and $\delta t = \frac{T}{M}$ with a large positive integer $M$.
The most naive way to approximate the operator $U(t)$ is
\begin{align}
\begin{split}
&U(t) 
=  e^{-\im H_{\rm ZZ}\delta t} e^{-\im H_{\pm,\mA}(t) \delta t} e^{-\im H_{Z,\mA}(t) \delta t}  
   +\mathcal{O}( \delta t ^2 ) .
\end{split}
\label{eq:ST}
\end{align}
While we use this approximation for $\theta =0$ with $(T,\delta t)=(100,0.1)$,
we use an improved version of \eqref{eq:ST} for $\theta \neq 0$
with $(T,\delta t)=(150,0.3)$ by using 
\begin{align}
U(t) 
&=  e^{-\im H_{\pm,\mA}(t) \frac{\delta t}{2}}e^{-\im H_{\rm ZZ}\delta t} e^{-\im H_{Z,\mA}(t) \delta t} e^{-\im H_{\pm,\mA}(t) \frac{\delta t}{2}}
\nonumber\\
& +\mathcal{O}( \delta t ^3 ),
\label{eq:ST2}
\end{align}
which is detailed in Supplementary Material.
We implement all the operators in the time evolution 
by combinations of quantum elementary gates provided by IBM Qiskit library
(see Supplementary Material).
Finally, in the process of measurement of the mass operator,
we take the number of shots to be $10^6$ in all the data points.
This induces statistical errors in addition to the systematic errors.

\subsection{Estimation of number of gates}

Here we have used a classical simulator for quantum hardware
to see how our algorithm practically works and grasp a future prospect
on applications of real quantum computers to quantum field theory.
The maximal number of qubits in our simulation is 16,
which is not so big even in current technology.
While this is quite encouraging,
the adiabatic preparation of state adopted here requires a large number of gates:
our quantum circuit for 16 qubits without improvement of Trotter decomposition 
has about 250 single-qubit gates and 270 two-qubit gates at each time step
which has been repeated about 1000 times.
This would make much noise and hard to perform stable simulations when we implement our simulation 
on NISQ devices.

\section{Results}
\label{sec:results}
\subsection{Massless case}
\begin{figure}[t]
\begin{center}
\includegraphics[width=63mm]{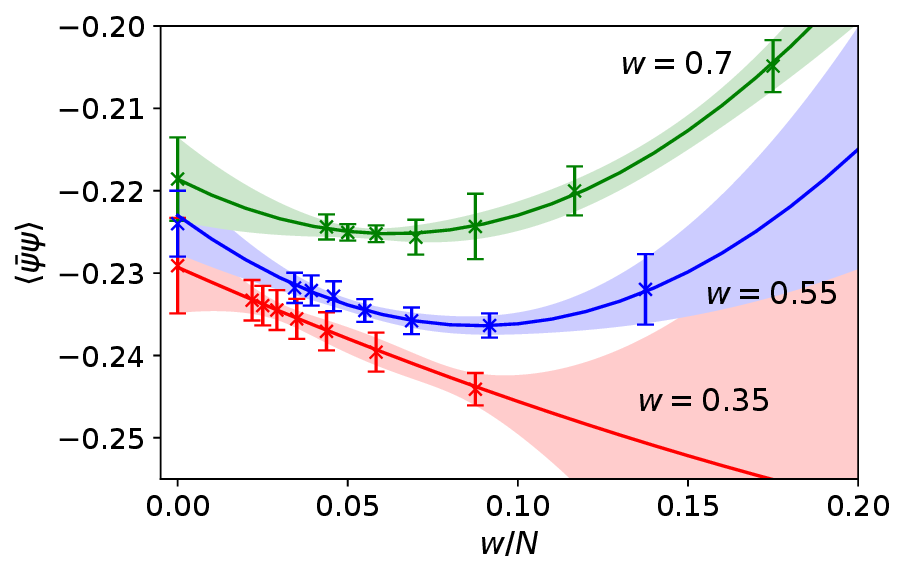}\\
\includegraphics[width=63mm]{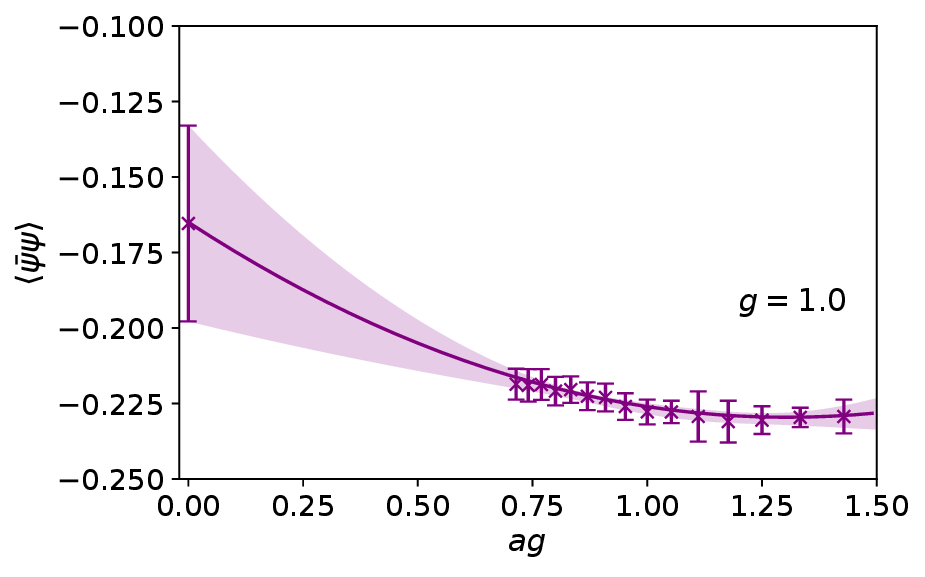}
\caption{
[Top]
Infinite volume limits for some values of $w$ at $g=1$, $m=0$, $\theta =0$.
For each $w$, we compute the VEV of the mass operator $\langle\bar\psi \psi\rangle$ for $N=[4,16]$ and extrapolate it to $N\rightarrow\infty$ by fitting the data to a quadratic polynomial of $w/N$ shown by the solid curves.
The error bars in the data points and error bands include both statistical 
and systematic errors.
[Bottom]
Continuum limits at $m=0, \theta =0$ for $g=1$.
The solid curve shows the fit function which is a quadratic polynomial in $ag$. 
The error bars and the error band take account of the extrapolation errors as well as the statistical and other systematic errors.
}
 \label{fig:continuum}
  \end{center}
\end{figure}

Let us first focus on the massless case and compare with the exact result
in the continuum theory 
\cite{Hetrick:1988yg,Hetrick:1995wq,Hosotani:1996hy,Hosotani:1996sn},
\begin{align}
\langle \bar{\psi}(x) \psi (x) \rangle 
= -\frac{e^\gamma }{2\pi^{3/2}} g
\approx -  0.160g ,
\label{eq:exact}
\end{align}
where $\gamma$ is the Euler-Mascheroni constant.
Note that the $\theta$-parameter is irrelevant in this case 
since our lattice Hamiltonian is independent of $\theta$ for $m=0$.
We take a physical limit for fixed physical parameters $(g,m,\theta )$ in two steps:
(i) Take infinite volume limit.
Namely, for fixed $w=1/2a$ and the physical parameters, 
we compute the observables for various $N$'s and 
then extrapolate them to $N\rightarrow\infty$ with quadratic polynomial in $1/N$.
Repeating this for multiple values of $w$, we obtain data of infinite volume limit for various lattice spacing $a$ at fixed physical parameters.
This step is illustrated in fig.~\ref{fig:continuum} [Left].
(ii) Extrapolate the data of the infinite volume limit 
to the continuum limit $a\rightarrow 0$ with quadratic polynomial in $ag$. 
This procedure is demonstrated in fig.~\ref{fig:continuum} [Right].


\begin{figure}[t]
\begin{center}
\includegraphics[width=70mm]{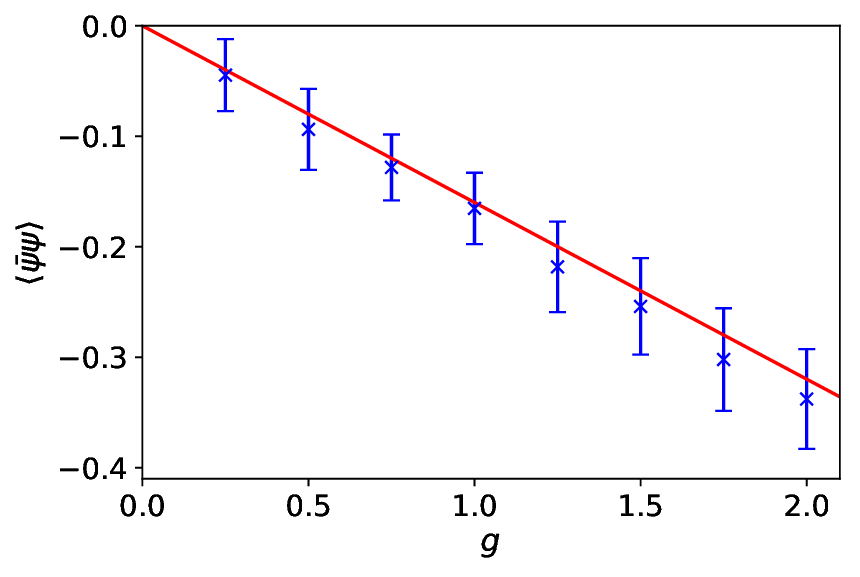}
\caption{
The VEV of the mass operator for $m=0$ is plotted against coupling constant $g$.
The red solid line shows the exact result, which is approximately $-0.160g$.
The error bars take account of extrapolation errors as well as the statistical and systematic errors.
}
\label{fig:massless}
  \end{center}
\end{figure}

Repeating the above procedures,
we have obtained $g$ dependence of the mass operator in the continuum limit
as shown in fig.~\ref{fig:massless}.
We see that
our result agrees with the exact result.
Note that the massless case cannot be easily explored by the standard Monte Carlo approach
because computational cost to evaluate effects of fermions in the standard approach is $O\big((am)^{-1}\big)$ \cite{Luscher:2010ae}.
This point is another advantage of our approach over the standard Monte Carlo approach.

\subsection{Massive case}
Next, we consider the massive case.
For this case, there is a result by mass perturbation theory \cite{Adam:1998tw,Adam:1997wt}:
\begin{align}
\langle \bar{\psi}(x) \psi (x) \rangle 
\approx  -0.160g +0.322m\cos{\theta} ,
\label{eq:perturbation}
\end{align}
up to $\mathcal{O}(m^2 )$.
There is a subtlety in comparison with this result:
the observable is UV divergent logarithmically and we need to regularize it.
Here we adopt a lattice counterpart of a regularization used in \cite{Adam:1998tw}
which is a subtraction of the free theory result.
Specifically, we take infinite volume limit without subtraction 
as in fig.~\ref{fig:continuum} [Left]
but subtract the result at $J=0$ and $N\rightarrow\infty $
in taking the continuum limit:
\begin{align}
&\langle \bar{\psi}(x) \psi (x) \rangle_{\rm free}
= -\frac{m\cos{\theta}}{\pi\sqrt{ 1+(ma \cos{\theta} )^2  }}
K(z) , \nonumber \\
& K(z)
=\int_0^{\frac{\pi}{2}} \frac{dt}{\sqrt{1-z\sin^2{t}}} ,
\quad
z=\frac{1 -(ma  \sin{\theta} )^2}{1+(ma \cos{\theta})^2 }  .
\end{align}
In other words, we replace $\langle \bar{\psi} \psi  \rangle$ in fig.~\ref{fig:continuum} [Right]
by $\langle \bar{\psi} \psi  \rangle$ $-$ $\langle \bar{\psi} \psi  \rangle_{\rm free}$
for the massive case.

\begin{figure}[t]
\begin{center}
\includegraphics[width=70mm]{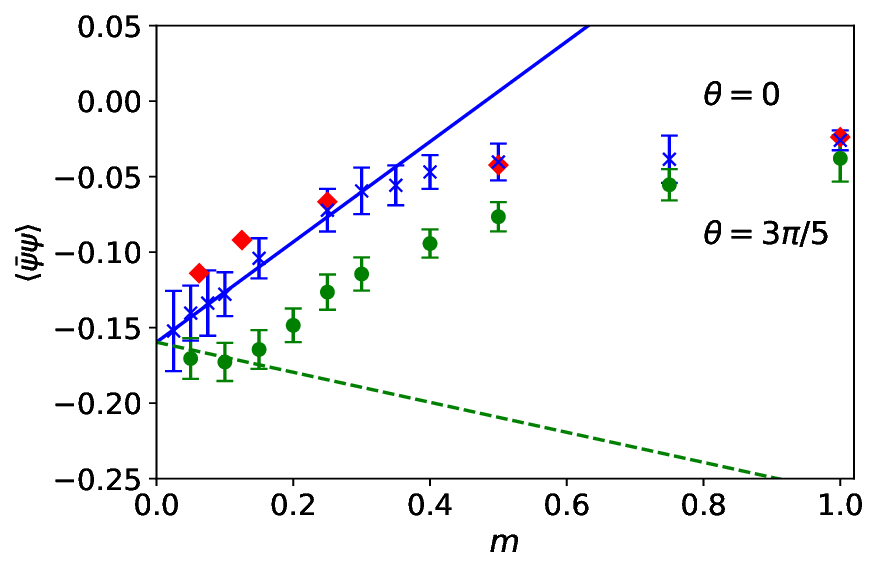}
\caption{
The VEV of the mass operator at $g=1$ is plotted against the mass $m$
for $\theta =0$ (blue $\times$'s) and $\theta =3\pi /5$ (green circles).
The red diamonds show the numerical result at $\theta =0$
obtained by the tensor network approach in \cite{Banuls:2016lkq}.
The lines show the result \eqref{eq:perturbation} of the mass perturbation.
The error bars are fitting errors in taking the continuum limit.
}
\label{fig:massive}
  \end{center}
\end{figure}

\begin{figure}[t]
\begin{center}
\includegraphics[width=70mm]{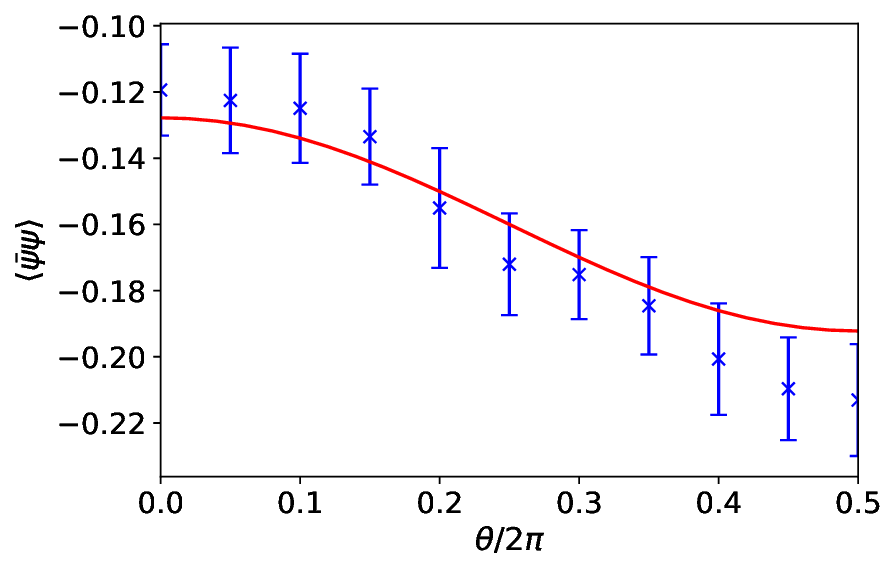}
\caption{
The VEV of the mass operator at $g=1$ and $m=0.1$ is plotted against $\theta /2\pi$ (blue symbols).
The curve shows the result \eqref{eq:perturbation} of the mass perturbation with fixed $g$ and $m$.
}
\label{fig:theta}
  \end{center}
\end{figure}

In fig.~\ref{fig:massive}, 
we plot our result in the continuum limit  
for $(g,\theta ) =(1,0)$ and $(g,\theta ) =(1,3\pi /5 )$ against $m$,
and compare with the mass perturbation theory \eqref{eq:perturbation}.
We see that 
our result agrees with the mass perturbation theory in small mass regime
for the both values of $\theta$.
As increasing mass,
it deviates from \eqref{eq:perturbation} and finally approaches zero.
This large mass asymptotic behavior is expected
since the large mass limit should be the same as the free theory result
which we have subtracted.
Our result for $\theta =0$ also agrees 
with previous numerical result obtained by tensor network approach \cite{Banuls:2016lkq}.
Furthermore, we compare $\theta$-dependence of the VEV of the mass operator for the parameters in Fig.~\ref{fig:theta},
for which the mass perturbation \eqref{eq:perturbation} is expected to be well behaved: $(g,m)=(1.0,0.1)$\footnote{
It is expected that the Schwinger model at $\theta =\pi$ has a critical point at $m\simeq 0.33g$ 
and first order phase transition associated with spontaneous breaking of charge conjugation symmetry (or equivalently parity) for larger $m$.
Therefore we are not passing any phase transition point in fig.~\ref{fig:theta}. 
}.
We see that our data agrees with the mass perturbation for most values of $\theta$
while we get small deviations around $\theta =\pi$ within our current accuracy.
Thus we conclude that
our approach practically works well for nonzero $(g,m,\theta )$.

\section{Summary and Discussions }
\label{sec:discussions}
In this paper,
we have designed the protocol for the digital quantum simulation of the Schwinger model with the $\theta$-term
as the first example of applications of quantum algorithms for gauge theory with topological terms.
We have converted the Schwinger model to the spin system on the spatial lattice
and then constructed the true vacuum state of the model using adiabatic state preparation.
We have computed VEV of the fermion mass operator,
taken the continuum limit and found agreement with the results in literature.
Our results imply that 
digital quantum simulation is already useful tool to explore 
non-perturbative aspects of gauge theories with real time and topological terms.

Our estimation for number of gates in our simulation hints at large noise and difficulty performing stable simulations when implemented on NISQ devices. Therefore it is important to save the number of gates and
reduce noise by improving the algorithm.
One way is to improve the Suzuki-Trotter decomposition
so that we can take smaller time steps $M$ to achieve similar errors.
Another way is to change the adiabatic Hamiltonian \eqref{eq:adiabaticH}
so that we can take smaller adiabatic time $T$.
In principle, the adiabatic Hamiltonian $H_A(t)$ can be any hermitian operator
satisfying \eqref{eq:adiabatic_cond} 
as long as the system during the adiabatic process is gapped and has a unique ground state.

There are various interesting applications and generalization of our work.
An obvious application is to compute other observables in the Schwinger model.
Specifically, the massive Schwinger model is known to exhibit confinement \cite{Gross:1995bp}
and therefore it would be interesting to explore physics of confinement
in a situation where standard Monte Carlo approach is inapplicable.
Another interesting direction is to apply our methods to other theories.
Our formulation to rewrite gauge theory in terms of qubits
can be directly applied
to any $1+1$ dimensional $U(1)$ gauge theory coupled to fermions.
It would also be interesting to implement a digital quantum simulation 
of $1+1$ dimensional non-Abelian gauge theories.

\subsection*{Acknowledgement}
The results of this work have been obtained by using quantum simulator in IBM Qiskit library.
Y.K. would like to thank Dmitri E. Kharzeev for discussions on closely related projects.
B.C. and M.H. have been partially supported by STFC consolidated grant ST/P000681/1. B.C. has been also supported by the Isaac Newton Trust, the
Leverhulme Trust ECF scheme.
The work of A.T. was supported by the RIKEN Special Postdoctoral Researcher program.
T.I. is supported in part by
US DOE Contract DESC0012704(BNL). T.I. is also supported by JSPS KAKENHI grant numbers JP26400261,
JP17H02906.

\appendix
\section{Operator correspondence between Dirac and Staggered fermions}
\label{app:fermion}
The Dirac fermion operator $\psi(x)=\big(\psi_u(x),\psi_d(x) \big)^\mathsf{T}$ 
is translated into those of staggered fermions as follows~\cite{Kogut:1974ag,Susskind:1976jm},
\begin{align}
\frac{\chi_n}{\sqrt{a}} \leftrightarrow \left\{ \begin{matrix} \psi_u (x) & n:{\rm even} \cr  
\psi_d (x) & n:{\rm odd} \end{matrix}\right. .
\end{align}
Here, we see how the bilinear operators $\bar{\psi}\gamma^1\p_1\psi$, $\bar{\psi}\psi$, and $\bar{\psi}\gamma^5\psi$ are written in terms of the staggered fermions.
We start with the fermion kinetic operator $\bar{\psi}\gamma^1\p_1\psi$:
\begin{align}
\begin{split}
&\bar{\psi}(x)\gamma^1\p_1\psi(x) 
\\
&= \psi^\dag_{u}(x)\frac{\psi_{d}(x+1)-\psi_{d}(x)}{2a}+\psi^\dag_{d}(x)\frac{\psi_{u}(x)-\psi_{u}(x-1)}{2a},
\\
&= \frac{1}{2a^2}\left[\chi^\dag_{2x}(\chi_{2x+1}-\chi_{2x-1})+\chi^\dag_{2x-1}(\chi_{2x}-\chi_{2x-2})\right].
\end{split}
\end{align}
Thus, we arrive at the following expression,
\begin{align}
a\sum_{x=1}^{N/2}\bar{\psi}(x)\gamma^1\p_1\psi(x)
=\frac{1}{2a}\sum_{n=1}^{N-1}\big[\chi^\dag_{n}\chi_{n+1}-\chi^\dag_{n+1}\chi_{n}\big].
\end{align}
Next, we convert the fermion mass operator $\bar{\psi}\psi$:
\begin{align}
\begin{split}
\bar{\psi}(x)\psi(x) &= \psi^\dag_{u}(x)\psi_{u}(x)-\psi^\dag_{d}(x)\psi_{d}(x),
\\
&=\frac{1}{a}\left[\chi^\dag_{2x}\chi_{2x}-\chi^\dag_{2x-1}\chi_{2x-1}\right],
\end{split}
\end{align}
which leads us to 
\begin{align}
a\sum_{x=1}^{N/2}\bar{\psi}(x)\psi(x)
=\sum_{n=1}^{N}(-1)^n\chi^\dag_{n}\chi_{n}.
\end{align}
Finally, we consider the pseudo mass operator $\bar{\psi}\gamma_5\psi$. 
Since it is a fermion bilinear operator involving off-diagonal matrix, the conversion to staggered fermion yields hopping term connecting even and odd sites:
\begin{align}
\begin{split}
&\bar{\psi}(x)\gamma_5\psi(x) = \psi^\dag_{u}(x)\psi_{d}(x)-\psi^\dag_{d}(x)\psi_{u}(x)
\\
&\approx \frac{1}{2}\left[\psi^\dag_{u}(x)\psi_{d}(x)-\psi^\dag_{d}(x)\psi_{u}(x)\right]
\\
&+\frac{1}{2}\left[\psi^\dag_{u}(x)\psi_{d}(x+1)-\psi^\dag_{d}(x+1)\psi_{u}(x)\right]
\\
&=-\frac{1}{2a}\left[\chi^\dag_{2x-1}\chi_{2x}-\chi^\dag_{2x}\chi_{2x-1}\right]
\\
&+\frac{1}{2a}\left[\chi^\dag_{2x}\chi_{2x+1}-\chi^\dag_{2x+1}\chi_{2x}\right],
\end{split}
\end{align}
where we have used $\psi_{d}(x+1) = \psi_{d}(x)+\calO(a)$ to modify the operator, that recovers the original one in the continuum limit $a\to0$.
Thus the pseudo mass operator is rewritten as
\begin{align}
\sum_{x=1}^{N/2}\bar{\psi}(x)\gamma_5\psi(x)
=\frac{1}{2} \sum_{n=1}^{N-1}(-1)^n\left[\chi^\dag_{n}\chi_{n+1}-\chi^\dag_{n+1}\chi_{n}\right].
\end{align}

\section{Alternative method: without chiral rotation}
\label{app:naive}
Here we rewrite the Schwinger model based on the Lagrangian (1)
without the chiral rotation in terms of the spin variables.
In the temporal gauge, conjugate momentum of $A^1$ is 
\begin{align}
\Pi = \frac{\del \cal{L}}{\del \dot{A}^1} = \dot{A}^1 +\frac{g\theta}{2\pi} ,
\end{align} 
and therefore the Hamiltonian is
\begin{align}
H
=\int dx \Biggl[ -\im\bar{\psi} \gamma^1 ( \del_1 +\im gA_1 )\psi  +m\bar{\psi}\psi 
+\frac{1}{2}\left( \Pi -\frac{g\theta}{2\pi} \right)^2
\Biggr].
\end{align}
Using the staggered fermion, the lattice Hamiltonian is given by
\begin{align}
H
&= -iw \sum_{n=1}^{N-1} \Bigl[ \chi_n^\dag e^{\im\phi_n} \psi_{n+1} -\chi_{n+1}^\dag e^{-\im\phi_n} \chi_{n}\Bigr] \br
  +m\sum_{n=1}^N (-1)^n \chi_n^\dag \chi_n 
+J\sum_{n=1}^{N-1} \left( L_n +\frac{\theta}{2\pi} \right)^2 ,
\end{align}
where $L_n$ corresponds to $-\Pi (x)/g$ and the operators satisfy the commutation relations (7)
as well as the Gauss law (8)
on physical states.
Applying the Jordan-Wigner transformation (9),
taking the open boundary condition with constant $L_0$ and solving the Gauss law,
we obtain the lattice Hamiltonian
\begin{align}
H
&= w \sum_{n=1}^{N-1} \Bigl[ \sigma_n^+  \sigma_{n+1}^-  +{\rm h.c.} \Bigr] 
  +\frac{m}{2}\sum_{n=1}^N (-1)^n  Z_n  \br
  +J \sum_{n=1}^{N-1} \Bigl[ 
    L_0 +\frac{\theta}{2\pi} + \frac{1}{2}\sum_{\ell =1}^n \left( Z_\ell +(-1)^\ell \right) \Bigr]^2 .
\end{align}
In this formulation, it is clear that
the theory with $(\theta ,L_0)$ is equivalent to $(\theta +2\pi L_0 ,0)$.
Digital quantum simulation in this formulation can be implemented in a similar way to the formulation in the main text.
It would be interesting to perform a digital quantum simulation in this formulation
and see how this formulation practically works.

\section{Details on quantum simulation protocol}
\label{app:details}
Here we write down all the qubit operations
used in quantum circuits in this paper.
First we write down single qubit operation which acts on a superposition of
\begin{align}
|0\rangle =\begin{pmatrix} 1 \cr 0 \end{pmatrix} \quad {\rm and}\quad
|1\rangle =\begin{pmatrix} 0 \cr 1 \end{pmatrix} .
\end{align}
Some of most basic operations are Pauli matrices:
\begin{align}
X= \begin{pmatrix} 0 & 1 \cr 1 & 0 \end{pmatrix} ,\quad
Y= \begin{pmatrix} 0 & -\im \cr \im & 0 \end{pmatrix} ,\quad
Z= \begin{pmatrix} 1 & 0 \cr 0 & -1 \end{pmatrix}
\end{align}
In terms of $(X,Y,Z)$, we also use
\begin{align}
R_X (\phi )
=e^{-\frac{\im\phi}{2}X} ,\quad
R_Y (\phi )
=e^{-\frac{\im\phi}{2}Y} ,\quad
R_Z (\phi )
=e^{-\frac{\im\phi}{2}Z} .
\end{align}
The only two qubit operation used in this paper is controlled-$X$ (controlled-NOT):

\begin{equation}
CX= \begin{pmatrix} 
1 & 0  & 0 & 0 \cr  
0 & 1  & 0 & 0 \cr  
0 & 0  & 0 & 1 \cr  
0 & 0  & 1 & 0 \cr  
\end{pmatrix} =
\begin{array}{c}
\Qcircuit @C=1em @R=.7em {
& \ctrl{1} & \qw
\\
& \targ{} & \qw
}
\end{array}
\end{equation}

which acts on superposition of $|i\rangle \otimes |j\rangle$ with $i,j=0,1$.
In particular, $CX$ satisfies
\begin{align}
CX |0\rangle \otimes |\alpha\rangle =  |0\rangle \otimes  |\alpha\rangle ,\quad
CX |1\rangle \otimes |\alpha\rangle =  |1\rangle \otimes  X |\alpha\rangle .
\end{align}

We can construct all the operators in (23)
by combinations of the quantum elementary gates $R_{X,Y,Z}$ and $CX$.
First, $e^{-\im H_Z \delta t}$ is simply realized by a product of single qubit operations:
\begin{align}
e^{-\im H_Z \delta t} 
=\prod_{n=1}^N R_Z^{(n)} (2c_n \delta t ) ,
\end{align}
where $R_Z^{(n)}(\phi )$ stands for a $R_Z (\phi )$ gate acting on $n$-th qubit 
and $c_n$ is defined by $\sum_{n=1}^N c_n Z_n = H_{Z,\mA}(t)$.
The other two unitary operators in (23)
involve two-qubit operations.
The operator $e^{-\im H_{\rm ZZ}\delta t}$ needs the following two-qubit operations of
\begin{align}
e^{-\im \frac{J\delta t}{2}Z_1 Z_2} ,
\end{align}
to appropriate pairs of qubits.
This operator is the same as the interaction of the Ising model
and its concrete realization is, 
\begin{align}
e^{-\im \frac{J\delta t}{2}Z_1 Z_2} =CX^{(12)}   R_Z^{(2)} (J \delta t )  CX^{(12)} ,
\end{align}
with a quantum gate given by
\begin{align}
\begin{split}
e^{-\im \frac{J\delta t}{2}Z_1 Z_2}=
\begin{array}{c}
\Qcircuit @C=1em @R=.7em {
& \multigate{1}{Z_1Z_2(\frac{J\delta t}{2})} &\qw 
\\
& \ghost{Z_1Z_2(\frac{J\delta t}{2})}& \qw
}
\end{array}
\\
:=
\begin{array}{c}
\Qcircuit @C=1em @R=.7em {
& \ctrl{1} & \qw & \ctrl{1} & \qw 
\\
& \targ{} & \gate{R_Z(J\delta t)} & \targ{} & \qw
}
\end{array}
\end{split}
\end{align}
The operator $e^{-\im H_\pm \frac{t\delta t}{T}}$ can be constructed in a similar way.
It needs the two qubit operations of
\begin{align}
e^{-\im \frac{\tilde{w}\delta t}{2} (X_1 X_2 +Y_1 Y_2)} 
=e^{-\im \frac{\tilde{w}\delta t}{2} X_1 X_2}e^{-\im \frac{\tilde{w}\delta t}{2}Y_1 Y_2} + \calO(\delta t^2) ,
\end{align}
to appropriate pairs. Here, $\tilde{w}$ is defined 
by $\tilde{w}:=\frac{t}{T}w 
-\frac{(-1)^n}{2}\left( \left(1-\frac{t}{T}\right) m_0 +m \right)\sin\big(\theta\frac{t}{T}\big)$.
This is concretely realized by
\begin{align}
& e^{-\im \frac{\tilde{w}\delta t}{2} X_1 X_2}
=  CX^{(12)} R_X^{(1)} ( \tilde{w}\delta t ) CX^{(12)}  ,\\
& e^{-\im \frac{\tilde{w}\delta t}{2} Y_1 Y_2}
=  \prod_{j=1}^2  R_Z^{(j)}\left( \frac{\pi}{2}\right) \cdot
e^{-\im \frac{\tilde{w}\delta t}{2} X_1 X_2} \cdot
 \prod_{j=1}^2  R_Z^{(j)}\left( -\frac{\pi}{2}\right)  ,
\end{align}
whose circuit diagrams are respectively given by
\begin{align}
\begin{split}
e^{-\im \frac{\tilde{w}\delta t}{2} X_1 X_2}=
\begin{array}{c}
\Qcircuit @C=1em @R=.7em {
 & \multigate{1}{X_1X_2(\frac{\tilde{w}\delta t}{2})} &\qw 
\\
 & \ghost{X_1X_2(\frac{\tilde{w}\delta t}{2})}& \qw
}
\end{array}
\\
:=
\begin{array}{c} 
\Qcircuit @C=1em @R=.7em {
 & \targ{} & \qw & \targ{} & \qw 
\\
 & \ctrl{-1} & \gate{R_X(\tilde{w}\delta t)} & \ctrl{-1} & \qw
}
\end{array}
\end{split}
\\[4mm]
%
\begin{split}
&e^{-\im \frac{\tilde{w}\delta t}{2}Y_1 Y_2} =
\begin{array}{c}
\Qcircuit @C=1em @R=.7em {
& \multigate{1}{Y_1Y_2(\frac{\tilde{w}\delta t}{2})} &\qw 
\\
& \ghost{Y_1Y_2(\frac{\tilde{w}\delta t}{2})}& \qw
}
\end{array}
\\[2mm]
&:=
\begin{array}{c}
\Qcircuit @C=1em @R=.7em {
&\gate{R_Z(-\frac{\pi}{2})}& \multigate{1}{X_1X_2(\frac{\tilde{w}\delta t}{2})} &\gate{R_Z(\frac{\pi}{2})}& \qw 
\\
&\gate{R_Z(-\frac{\pi}{2})}& \ghost{X_1X_2(\frac{\tilde{w}\delta t}{2})} &\gate{R_Z(\frac{\pi}{2})}& \qw
}
\end{array}
\end{split}
\end{align}

For example,
we implement the time evolution operator $U(t)$~(23)
with lattice size $N=4$ by the following quantum circuit:
\begin{widetext}
\begin{align}
\begin{array}{c}
\Qcircuit @C=1em @R=.7em {
\lstick{n=4} &\gate{R_Z^{(4)}} & \multigate{1}{Y_3Y_4}&\qw& \multigate{1}{X_3X_4}&\qw & \qw &\qw&\qw &\qw
\\
\lstick{n=3} &\gate{R_Z^{(3)}} & \ghost{Y_3Y_4}&\multigate{1}{Y_2Y_3}&\ghost{X_3X_4} &\multigate{1}{X_2X_3}      &\qw    &\multigate{2}{\text{\parbox{0.8cm}{$Z_1Z_3$\\ }}}& \multigate{1}{Z_2Z_3}&\qw 
\\
\lstick{n=2} &\gate{R_Z^{(2)}} & \multigate{1}{Y_1Y_2}&\ghost{Y_2Y_3}& \multigate{1}{X_1X_2}& \ghost{X_2X_3}     & \multigate{1}{Z_1Z_2}&\qw&\ghost{Z_1Z_3}  &\qw 
\\
\lstick{n=1} &\gate{R_Z^{(1)}} &\ghost{Y_1Y_2} &\qw &\ghost{X_1X_2} &\qw                                   &\ghost{Z_1Z_3} & \ghost{Z_1Z_3}&\qw &\qw 
}
\end{array}
\end{align}
\end{widetext}
where the argument of each unitary gate is suppressed and can be read off from (12), (21), and (23):
$R_Z^{(n)} \to R^{(n)}_Z(2c_n\delta t)$,
$X_{n}X_{n+1} \to X_{n}X_{n+1}\big(\frac{\tilde{w}\delta t}{2}\big)$, 
$Y_{n}Y_{n+1} \to Y_{n}Y_{n+1}\big(\frac{\tilde{w}\delta t}{2}\big)$,
$Z_{1}Z_{2} \to Z_{1}Z_{2}(J\delta t)$, 
$Z_{1}Z_{3} \to Z_{1}Z_{3}\big(\frac{J\delta t}{2}\big)$ and
$Z_{2}Z_{3} \to Z_{2}Z_{3}\big(\frac{J\delta t}{2}\big)$.

\section{Estimation of systematic errors}
\label{app:error}
Here we explain how we estimate systematic errors shown in the main text.
A VEV of an operator $\mathcal{O}$ is defined by
\begin{align}
\langle \mathcal{O} \rangle 
=\langle 0 | \mathcal{O} |0\rangle ,
\label{eq:VEV}
\end{align}
where here we denote ground state of a system under consideration by $|0\rangle$. 
Suppose we would like to find an approximation of this quantity 
by using an adiabatic preparation of the vacuum as in the main text.
Let us denote the approximate vacuum obtained in this way as $|0_A \rangle$.
Then we approximate the VEV \eqref{eq:VEV} by
\begin{align}
\langle \mathcal{O} \rangle_A 
=\langle 0_A | \mathcal{O} |0_A \rangle  ,
\label{eq:VEVapprox}
\end{align}
which is generically different from the true VEV.
The state $|0_A \rangle$ can be expanded as
\begin{align}
|0_A \rangle
=\sum_{n=0}^\infty c_n |n\rangle ,
\label{eq:expansion}
\end{align}
where $|n\rangle $ is the $n$-th excited state of the full Hamiltonian $H$ of the system.
If we take the adiabatic time $T$ and the number of steps $M$ 
in the Suzuki-Trotter decomposition sufficiently large, 
then we expect $|c_0|\simeq 1  \gg |c_{n\neq 0}|$
and $|0_A \rangle$ is almost the true vacuum.

Now we propose how to estimate systematic error 
in approximating the VEV \eqref{eq:VEV} by \eqref{eq:VEVapprox}.
Let us consider the quantity
\begin{align}
\langle \mathcal{O} \rangle_A (t) 
=\langle 0_A | e^{\im Ht} \mathcal{O} e^{-\im Ht} |0_A \rangle  .
\end{align}
If we managed to prepare the vacuum exactly i.e.~$|0\rangle_A =|0\rangle$,
then this quantity was reduced to $\langle \mathcal{O} \rangle $ and independent of $t$
since the vacuum is the eigenstate of $H$.
However, this quantity depends on $t$ 
when we have only approximation of the vacuum.
Let us see how it depends on $t$ using the expansion \eqref{eq:expansion}:
\begin{align}
\langle \mathcal{O} \rangle_A (t) 
&= \sum_{n=0}^\infty  |c_n |^2  \langle n |  \mathcal{O} |n \rangle  \NN\\
& +2\sum_{m\neq  n} {\rm Re} \left( c_m^\ast c_n  e^{\im(E_m -E_n )t} \langle m |  \mathcal{O} |n \rangle \right) ,
\end{align}
which implies that
this quantity oscillates 
around the constant $\sum_{n=0}^\infty  |c_n |^2  \langle n |  \mathcal{O} |n \rangle$
as varying $t$.
If we have a nice approximation of the vacuum s.t. $|c_0| \gg |c_{n\neq 0}|$,
then we approximately have
\begin{align}
\langle \mathcal{O} \rangle_A (t) 
&\simeq |c_0 |^2 \Biggl[  
\langle \mathcal{O} \rangle 
+\sum_{n=1}^\infty 
{\rm Re} \left( \frac{2c_n^\ast c_0}{|c_0 |^2}  e^{\im(E_n -E_0 )t} \langle n |  \mathcal{O} |0 \rangle \right) \br
 +\mathcal{O}\left( \left| \frac{c_n }{c_0} \right|^2 \right) \Biggr] ,
\end{align}
which approximately oscillates 
around $\simeq $ $\langle \mathcal{O} \rangle $.
Therefore the quantity $\langle \mathcal{O} \rangle_A (t) $ represents
intrinsic ambiguity in predicting the true VEV $\langle \mathcal{O} \rangle $
by the adiabatic state preparation.
Thus, in the main text,
we regard 
\begin{align}
\frac{1}{2} \left(  {\rm max}\langle\mathcal{O}\rangle_A (t) +{\rm min}\langle\mathcal{O}\rangle_A (t)  \right) 
\end{align}
as central value, and
\begin{align}
\frac{1}{2} \left(  {\rm max}\langle\mathcal{O}\rangle_A (t) -{\rm min}\langle\mathcal{O}\rangle_A (t)  \right) 
\end{align}
as systematic error in approximating the true VEV $\langle \mathcal{O} \rangle $ by the adiabatic preparation of the vacuum.
\begin{figure}[t]
\begin{center}
\includegraphics[width=42mm]{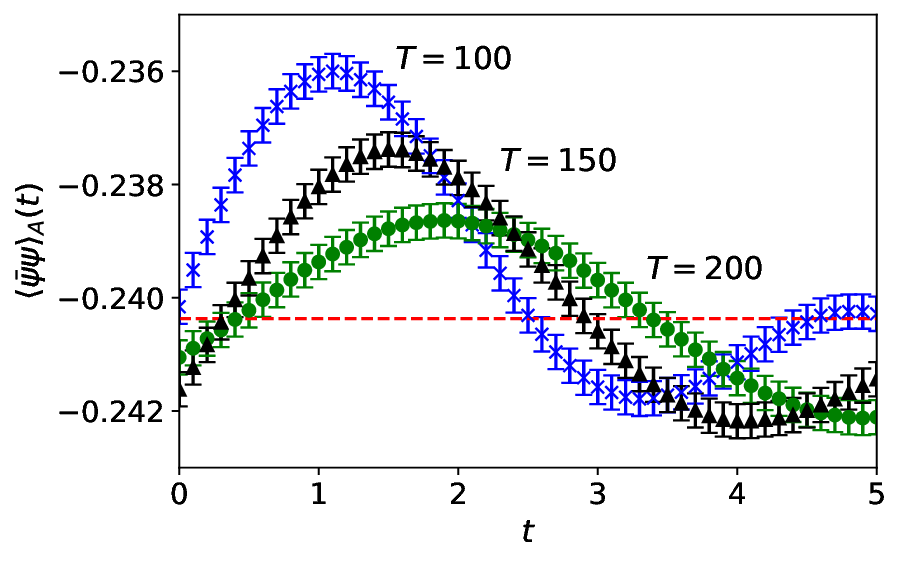}
\includegraphics[width=42mm]{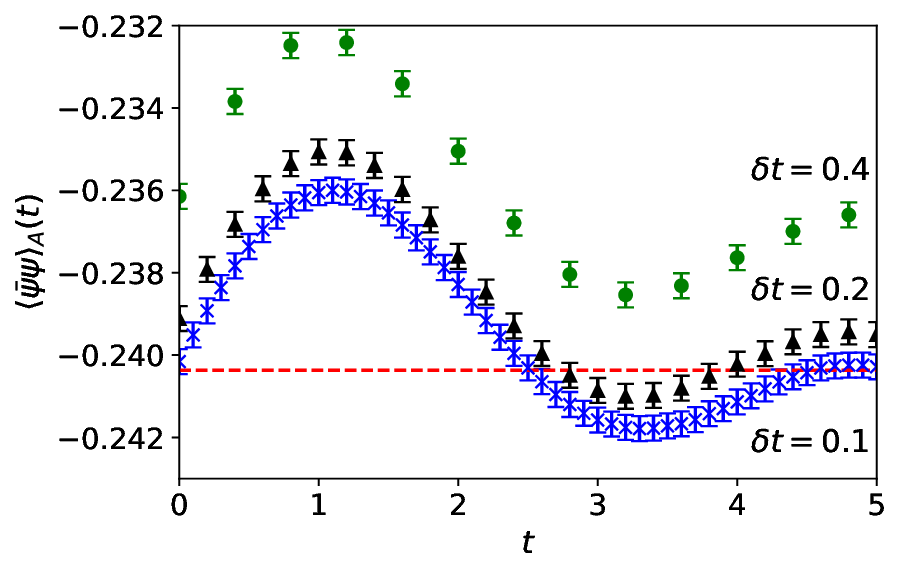}
\caption{
The expectation value of the mass operator under the state $e^{-\im Ht}|0_A \rangle$
for $(g,m,N,w)$ $=$ $(1,0,4,0.5)$ against $t$
obtained by simulations with $m_0 =0.5$ and $10^6$ shots.
The red dashed line is the result obtained by diagonalization of the Hamiltonian.
The error bars are statistical errors.
[Left] At fixed $\delta t=0.1$ with some values of $T$.
[Right] At fixed $T=100$ with some values of $\delta t$
}
\label{fig:error}
  \end{center}
\end{figure}
\begin{figure}[t]
\begin{center}
\includegraphics[width=50mm]{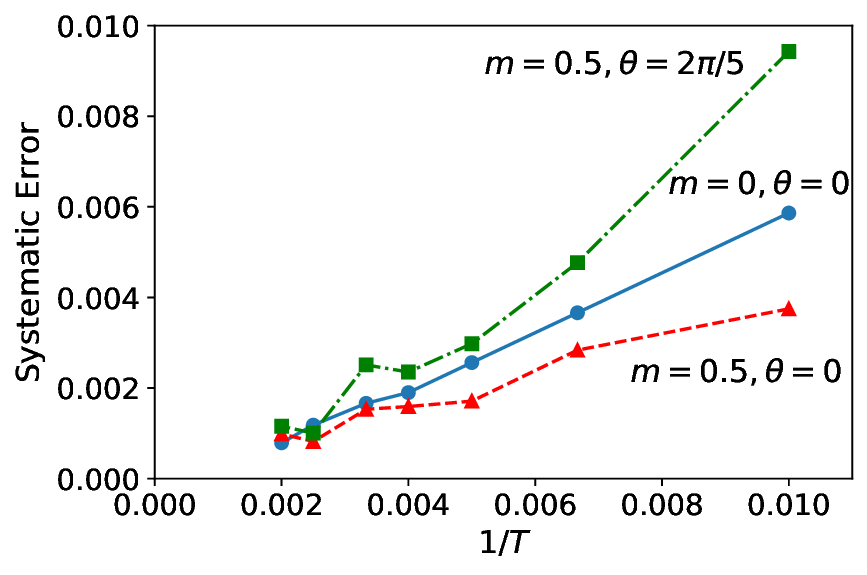}
\caption{
The systematic errors of the mass operator under the state $e^{-\im Ht}|0_A \rangle$
for $(g,N,w)$ $=$ $(1,8,0.5)$ against the inverse adiabatic time $1/T$
obtained by simulations with $10^6$ shots.
All the data points are obtained by exact time-evolution of the Hamiltonian without the Trotter decomposition.}
\label{fig:error_Tdep}
  \end{center}
\end{figure}

Fig.~\ref{fig:error} demonstrates the above procedure for the VEV of the mass operator
computed in the main text.
In fig.~\ref{fig:error} [Left],
we fix the Trotter step to $\delta t=0.1$ and
plot the results for different values of the adiabatic time $T$. 
We find that
the expectation value of the mass operator under the state $e^{-\im Ht}|0_A \rangle$
oscillates around the true VEV obtained of the Hamiltonian as expected.
We also find that
the result with larger $T$ has smaller amplitude.
This reflects the fact that
the approximate vacuum $|0_A \rangle$ with larger $T$ is closer to the true vacuum
and therefore the systematic error must be smaller for larger $T$.
In fig.~\ref{fig:error} [Right],
we fix $T$ as $T=100$ and
plot the results for different values of $\delta t$. 
The green circles show that
if we do not take sufficiently small $\delta t$,
then approximation of the time evolution operator $e^{-\im Ht}$ breaks down
and it does not oscillates around the correct value.
In fig.~\ref{fig:error_Tdep}, the adiabatic time dependence of the systematic errors associated 
with the mass operator is shown 
for different fermion masses and $\theta$ values.
All the curves roughly proportional to $1/T$. It nicely shows that the smaller fermion mass or larger $\theta$
result in larger systematic errors.
Thus, it is important to take appropriate values of $T$ and $\delta t$
to get reasonable approximations.

Finally, we comment on the potential reduction of the number of Trotter steps.
One can actually tell from Fig.~\ref{fig:error} that $10^{4}\sim10^5$ shots would be enough to maintain the same order of the total error. For instance, if we decrease the number of shots to $10^4$, the statistical error is roughly 10 times larger, whose magnitude is comparable to the adiabatic error.

\section{Improvement of the Suzuki-Trotter decomposition}
\label{app:Trotter}
The first-order Suzuki-Trotter decomposition is 
\begin{align}
 \e^{-\im(H_1+H_2)\delta t} = \e^{-\im H_1\delta t}\e^{-\im H_2\delta t} + \calO(\delta t^2),
\end{align}
for non-commuting operators $H_1$ and $H_2$. 
This error is reduced by using the second-order decomposition,
\begin{align}
 \e^{-\im(H_1+H_2)\delta t} = \e^{-\im H_1\frac{\delta t}{2}}\e^{-\im H_2\delta t}\e^{-\im H_1\frac{\delta t}{2}} + \calO(\delta t^3).
\end{align}
Let us apply this improvement to our adiabatic state preparation.
First we decompose the adiabatic Hamiltonian as
\begin{align}
H_A (t) =  \tilde{H}_Z(t) +\tilde{H}_X (t) +\tilde{H}_Y (t) ,
\end{align}
where
\begin{align}
\begin{split}
\tilde{H}_Z &= H_{ZZ} +H_{Z,\mA}(t) 
\\
\tilde{H}_X 
&=  \frac{1}{2}\sum_{n=1}^{N-1}h_{XY}(t)  X_nX_{n+1}   ,\\
\tilde{H}_Y 
&=  \frac{1}{2}\sum_{n=1}^{N-1}h_{XY}(t) Y_nY_{n+1} , \\
h_{XY}(t) &=\frac{t}{T}w 
-\frac{(-1)^n}{2}\left[\left( 1-\frac{t}{T} \right) m_0 +\frac{t}{T}m \right] 
\sin\left(\frac{t}{T}\theta\right).
\end{split}
\end{align}
This implies that the Hamiltonian can be divided into three sets of operators:
\begin{align}
\begin{split}
 \tilde{H}_{Z}&:  \{Z_1,\dots, Z_{N}, Z_1Z_2,Z_1Z_3,\dots,Z_{N-1}Z_N\}, \\
 \tilde{H}_{X}&: \{X_1X_2,X_2X_3,\dots, X_{N-1}X_{N}\}, \\
 \tilde{H}_{Y}&: \{Y_1Y_2,Y_2Y_3,\dots, Y_{N-1}Y_{N}\}.
\end{split}
\end{align}
The operators commute with each other within each set.
Then the time evolution operator $U(t)=e^{-\im H_A (t)\delta t}$ is approximated by
\begin{align}
U(t) 
=e^{-\im \tilde{H}_Y \frac{\delta t}{2}} e^{-\im \tilde{H}_X \frac{\delta t}{2}}    
  e^{-\im \tilde{H}_Z \delta t}    
 e^{-\im \tilde{H}_X \frac{\delta t}{2}}   e^{-\im \tilde{H}_Y \frac{\delta t}{2}}    
+\calO(\delta t^3).
\end{align}
In this improvement,
the quantum circuit for 16 qubits has about 400 single-qubit gates and 500 two-qubit gates at each time step
while the one without the improvement has
250 single-qubit gates and 270 two-qubit gates.
Note that 
the improvement saves the number of gates in the total time evolution 
since it needs smaller time steps to achieve the same accuracy.

\clearpage
\providecommand{\href}[2]{#2}\begingroup\raggedright\endgroup



\end{document}